\documentclass[twocolumn, prl]{revtex4}
\usepackage{graphicx}
\usepackage{epsfig}
\begin{document}
\title{Direct mapping of the finite temperature phase diagram of strongly correlated quantum models }
\author{Qi Zhou$^{1}$, Yasuyuki Kato$^{2}$, Naoki Kawashima$^{2}$, and Nandini Trivedi$^{1}$}
\affiliation{$^{1}$ Department of Physics, The Ohio State University, Columbus, OH 43210\\$^{2}$Institute for Solid State Physics, University of Tokyo, 5-1-5 Kashiwa-no-ha, Kashiwa, Chiba 277-8581, Japan }
\date{\today}
\begin{abstract}
Optical lattice experiments, with the unique potential of tuning interactions and density, have emerged as emulators of nontrivial theoretical models that are directly relevant for strongly correlated materials\cite{1,2,3,4, 5, 6, 7, 8, 9}. However, so far the finite temperature phase diagram has not been mapped out for any strongly correlated quantum model. We propose a remarkable method for obtaining such a phase diagram for the first time directly from experiments using only the density profile in the trap as the input. We illustrate the procedure explicitly for the Bose Hubbard model, a textbook example of a quantum phase transition from a superfluid to a Mott insulator\cite{10}. Using ``exactÓ quantum Monte Carlo simulations in a trap with up to $10^6$ bosons, we show that kinks in the local compressibility, arising from critical fluctuations, demarcate the boundaries between superfluid and normal phases in the trap. The temperature of the bosons in the optical lattice is determined from the density profile at the edge. Our method can be applied to other phase transitions even when reliable numerical results are not available. 
\end{abstract}
\maketitle

The grand challenge of condensed matter physics is to understand the emergence of novel phases arising from the organization of many degrees of freedom especially in regimes where particle interactions dominate over the kinetic energy. Brought to the forefront by the discovery of high temperature superconductivity in complex copper-based oxides, it is absolutely astounding that a simple model, like the Hubbard model is able to capture so many essential aspects of the physics of these materials. However, whether the repulsive fermion Hubbard model really contains a d-wave superconducting ground state is still an open question after decades of study.

Optical lattice experiments with cold atoms are emerging as an amazing laboratory for making realizations of such bose and fermi Hubbard and Heisenberg-type models and observing phase transitions without the uncertainty posed by the complex materials. In this backdrop quantum Monte Carlo simulations in strongly interacting regimes are emerging as an important bridge between materials based condensed matter physics and cold atoms, highlighted in Fig. 1. The scale of the numerical simulations possible today with up to $10^6$ particles is able to match the experimental cold atom systems thereby allowing a direct comparison.

\begin{figure}
\includegraphics[width=2.8in]{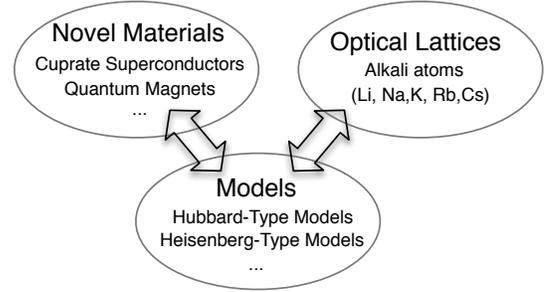}
\caption{Interaction between novel materials, models and optical lattice emulators. The models capture the fundamental physics of complex materials with novel phenomena such as high temperature superconductivity and quantum magnetism and form a bridge with experiments on ultracold atoms in optical lattices that emulate these models in a clean environment. }
\end{figure}

The most important task for the quantum emulator of a Hamiltonian $H$ is to determine the corresponding phase diagram in the temperature $T$ , chemical potential $\mu$ and {g} space, where the latter are parameters of $H$. So far such a mapping has not been determined experimentally in any model.  Even for the simplest Bose Hubbard model there have been many challenges arising from (i) lack of a clear diagnostic of how to identify phases. Traditional methods using the sharpness of the interference patterns are not reliable to distinguish the normal and superfluid phases; (ii) complications due to coexistence of different phases in the same confining potential; (iii) lack of thermometry of the Bose gas in the optical lattice; (iv) and until very recently, the lack of QMC simulations on experimentally relevant sizes. The work reported here overcomes all of the above obstacles. 

The Hamiltonian for the single band Bose Hubbard Model (BHM) is given by:
\begin{equation}
H_{BHM}=-\frac{t}{z}\sum_{\langle i,j\rangle}(b^\dagger_ib_j+h.c)+\frac{U}{2}\sum_in_i(n_i-1)-\mu_0\sum_in_i
\end{equation}                                          

\begin{figure*}
$\begin{array}{c}
\includegraphics[width=7in]{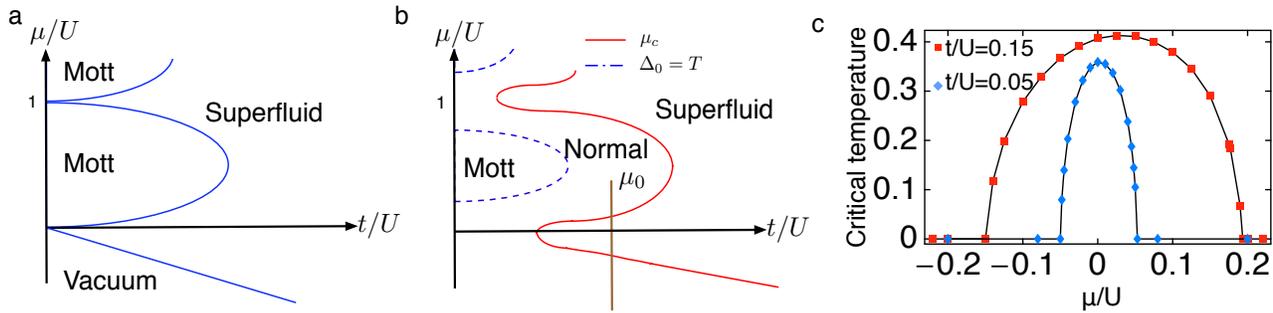}
\end{array}$
\caption{Phase diagram of BHM. (a) $\mu-t/U$ plane at $T=0$ (schematic); (b) $\mu-t/U$ plane at $T\neq 0$ (schematic). The superfluid phase is characterised by a non-zero order parameter $\langle b^\dagger\rangle \neq 0$ and a non-zero superfluid density $\rho_s\neq 0$, both of which vanish in the normal phase. The Mott insulator at $T=0$, characterized by integer filling and a finite gap to excitations $\Delta_0$, crosses over to the normal state for $T\approx\Delta_0$ . At finite temperatures, the critical chemical potential $\mu_c(T,t/U)$ demarcates the S-N phase boundary. In trapped atomic gases, the local chemical potential decreases from the centre of the trap to the edge (brown vertical line). (c) QMC simulations and finite size scaling to calculate the phase diagram in the $\mu-T$ plane for a fixed $t/U$.}
\end{figure*}
where $b_i$($b^\dagger_i$)  is the boson destruction (creation) operator at a site $i$,  $n_i=b^\dagger_ib_i$ is the density operator,  $z=6$ is the coordination number in 3D, and $\mu_0$  controls the density of bosons. The relative strength, $g=t/U$  of the tunnelling $t$ of bosons between nearest neighbor sites vs. the repulsive interaction $U$ between bosons, tunes the system through a superfluid to Mott transition at  $T=0$ (Fig. 2(a)). At finite $T$  the system shows a phase transition from a superfluid (S) to a normal(N) phase (Fig. 2(b)). In optical lattice experiments, bosons are confined in an additional harmonic trap of frequency $\omega$ modelled by $H=H_{BHM}-\frac{1}{2}m\omega^2\sum_ir^2_in_i$  where $m$ is the mass of the bosons. We simulate the Hamiltonian $H$  with the trap using quantum Monte Carlo techniques that include the effect of strong interactions ``exactly" within statistical errors. Recent modifications of the directed-loop algorithm for the world-line quantum Monte Carlo method\cite{11} have allowed us to significantly improve the efficiency near a critical point\cite{12} for large 3D systems with up to $10^6$ bosons in a $64^3$ lattice for the first time. Thus the phase diagram in the $\mu-T$ plane is calculated by finite size scaling for fixed $t/U$ .(Fig.2(c)).

Experimentally, the challenge of obtaining the phase diagram at finite $T$ lies in identifying measurable properties that can diagnose different phases. Traditionally the momentum distribution $n(\vec{k})$  imaged in the interference patterns of the expanding cold atom clouds has been used to identify the phases. In a long-time ballistic expansion the interference pattern $\tilde{n}(\vec{r})=\left({m}/{\hbar \tau}\right)^3|W\left(\vec{k}=\frac{m\vec{r}}{\hbar \tau}\right)|^2n\left(\vec{k}=\frac{m\vec{r}}{\hbar \tau}\right)$ essentially provides an image of $n(\vec{k})=\sum_{i,j}\langle b^\dagger_ib_j\rangle e^{i\vec{k}\cdot(\vec{r}_i-\vec{r}_j)}$ before expansion by convoluting with $W(\vec{k})$ , the Fourier transform of the Wannier function within a single site. Here $\tau$  is the expansion time. The final image detects the column integrated momentum distribution $N_\bot(k_x,k_y)=\int dk_z|W(\vec{k})|^2n(\vec{k})$. We have previously shown for a homogeneous BHM that sharp peaks in the interference pattern are not reliable for identifying a superfluid\cite{13,14,15}. What is the effect of a confining potential on $n(\vec{k})$? 
         
To answer this question we calculate both $n(\vec{k})$ and the density profile $\rho(r)$  in a harmonic trap at different $T$ and tuning $t/U$ (Fig. 3). For comparison, within local density approximation (LDA), we also calculate $\rho^h(\mu(r))$, the density $\rho^h(\mu)$  for a homogeneous system using QMC, where $\mu(r)=\mu_0-m\omega^2r^2/2$. Fig.3 shows the excellent agreement of $\rho(r)$ and $\rho^h(\mu(r))$, not unexpected for a shallow harmonic trap. We next turn to a key diagnostic, the local superfluid density distribution in the trap $\rho^h_s(\mu(r))$, obtained from a knowledge of $\mu(r)$ and the superfluid density $\rho^{h}_s(\mu)$ in the homogeneous system. The most significant observation from these data is that even in the presence of a trap when all the atoms are in the normal state, the interference pattern continues to show sharp peaks.  Strong repulsive interactions that expand the bosonic cloud and suppress $T_c$ in the optical lattice contribute to a sharpening up of $n(\vec{k})$ even in the normal state. With the lowering of $T$, as more regions within the trap become superfluid, there is a distinct change in the shape of the sharp peak (Fig. 3(d)). The emergence of a singular feature of width $\sim 1/L$, limited by the cloud size, indicates phase coherence throughout the sample. 
   
\begin{figure}[tp]
\includegraphics[width=3.5in]{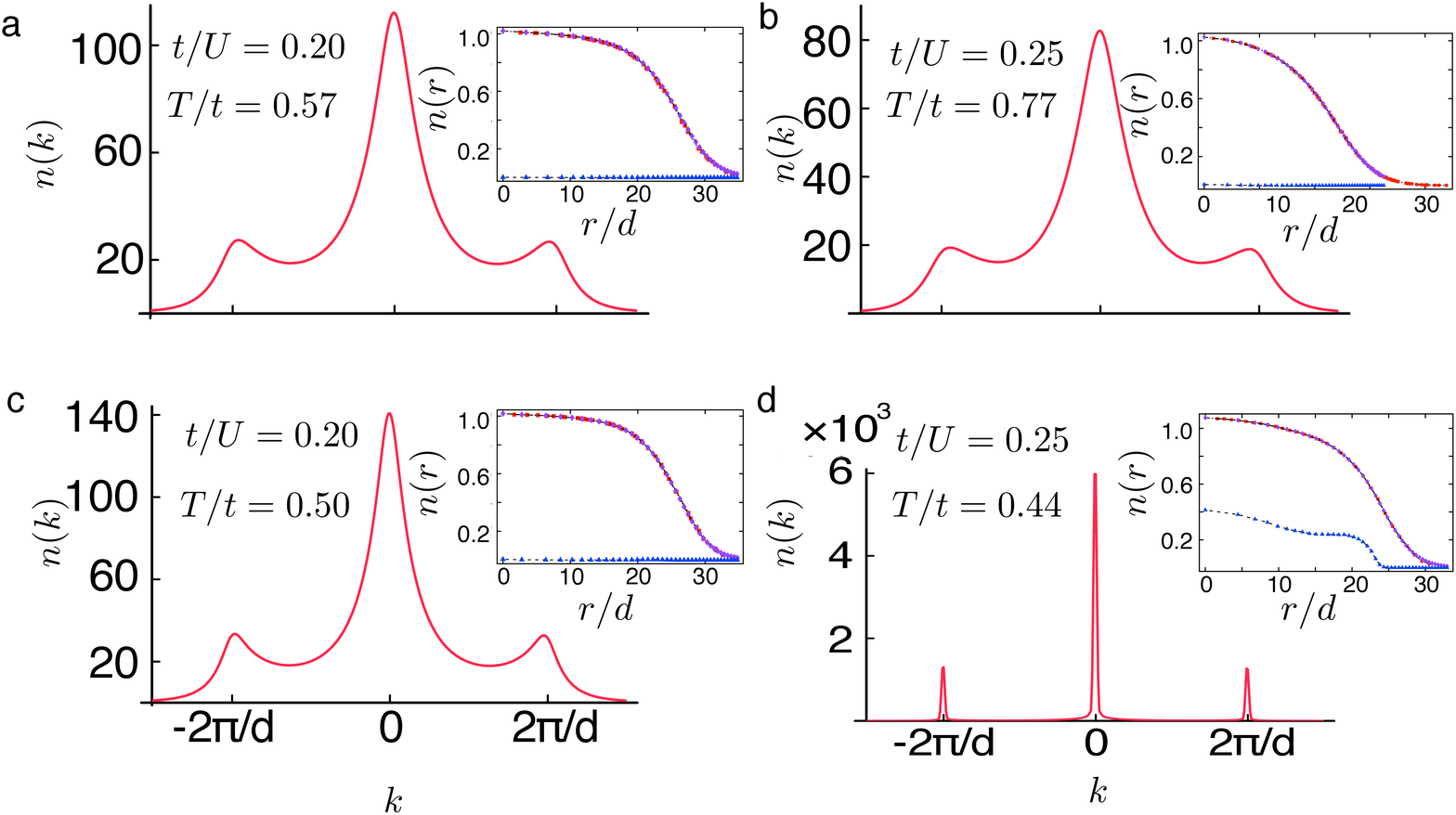}
\caption{Interference pattern in a trap with $N\approx10^5$ bosons: The insets show the local density profile $\rho(r)$(red boxes) obtained using QMC as a function of the radial coordinate in the trap. Also shown is the excellent agreement with $\rho^h(\mu(r))$ calculated within local density approximation (LDA) (purple diamonds). The superfluid density $\rho_s^h(\mu(r))$ is non-zero in a significant portion of the trap only at sufficiently low temperatures in panel (d). Note that sharp peaks are seen in $n(\vec{k})$ in panels (a), (b) and (c) even when all the atoms are in the normal state. The peak in panel (d) is much sharper. The difference between the normal state and the superfluid state is evident from the system-size dependence of the width of the peak.}
\end{figure}         
         
It is evident from Fig. 3 that $n(\vec{k})$ which integrates over the entire trap is unable to provide local information about the distribution of the phases. Also the finite expansion time, as well as resolution problems complicates the analysis of the interference pattern $\tilde{n}(\vec{r})$ and its relation to $n(\vec{k})$\cite{16}. We therefore explore direct methods of identifying phase boundaries in the trap. 
        
Our proposal relies on extracting the local compressibility in the trap defined by $\kappa_{diff}=\frac{-1}{m\omega^2r}\frac{d\rho}{dr}$ from a high-resolution scan of $\rho(r)$ in a trap. $\kappa_{diff}$ agrees very well with the compressibility $\kappa^h(\mu(r))=\frac{d\rho^h}{d\mu}$ for a bulk system (Fig.4(a)). A novel feature of the local compressibility is the existence of kinks at specific locations in the trap. The origin for the kink feature becomes evident if we consider in addition the behavior of the local superfluid density in the trap $\rho_s^h(\mu(r))$ (Fig.5). We see clearly that the kinks in the compressibility occur at the S-N phase boundary. At fixed $T$, the superfluid order becomes weaker when approaching the phase boundary. The increase of fluctuations is reflected in the growth of $\kappa^h(r)$. Starting from deep in the Mott-like region, where the particle number per site is essentially unity, $\kappa^h(r)$ increases when approaching the phase boundary with a superfluid resulting in a singular feature in $\kappa^h(r)$. The phase transition in the BHM is in the XY universality class with $\kappa^h(r)\sim |\mu-\mu_c|^{-\alpha}$  and $\alpha=-1$ giving a cusp at the transition. The singular feature is evident even with rounding from finite size effects in a trap as seen from Fig 4 and 5. At high temperatures, when $\rho_s$  vanishes in the trap, the kinks disappear in the absence of a phase boundary and the compressibility changes smoothly. 

\begin{figure}
\includegraphics[width=3.5in]{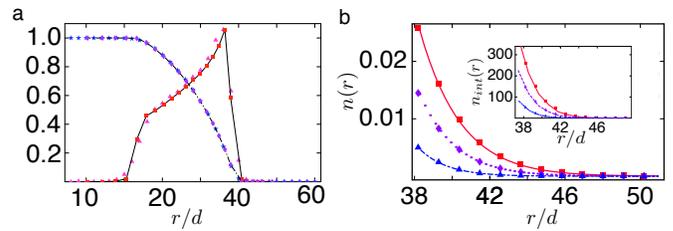}
\caption{Kinks in the local compressibility and mapping the phase diagram: (a) The local density $\rho(r)$(blue dots) agrees with $\rho^h(r)$(purple diamonds) calculated for a homogeneous system. We also show that the compressibility $\kappa_{diff}(r)$ (red boxes) directly obtained from the density profile agrees with $\kappa^h(\mu(r))$ (pink triangles) calculated within LDA. The signal to noise in the derivative method to extract $\kappa_{diff}(r)$ can be improved by taking an angular average of $\rho(r)$. Experimentally, $\rho(r)$ is obtained by an inverse Abel transformation of the column density\cite{20} or by identifying the planar-integrated density with the pressure. Recently, the column density has been measured in high resolution experiments\cite{21}. Alternatively,  $\rho(x,y,z_0)$ at fixed $z_0$ can be obtained directly\cite{22} using a ÒslidingÓ technique to get the density. (b) Determination of temperature by fitting (shown by lines) of the tail of the density profile (QMC results shown by symbols) to the ideal gas behavior in the dilute regime $e^{\beta\mu}\ll1$ and $|\mu(r)+2t|\gg Un$  so interactions can be ignored. Inset shows the fitting to the angular integrated densities at the corresponding temperatures that have larger signals and therefore better accuracy.  $T/t$ are $0.67$, $0.57$ and $0.44$ from top to bottom.}
\end{figure}         

\begin{figure}
\includegraphics[width=3.5in]{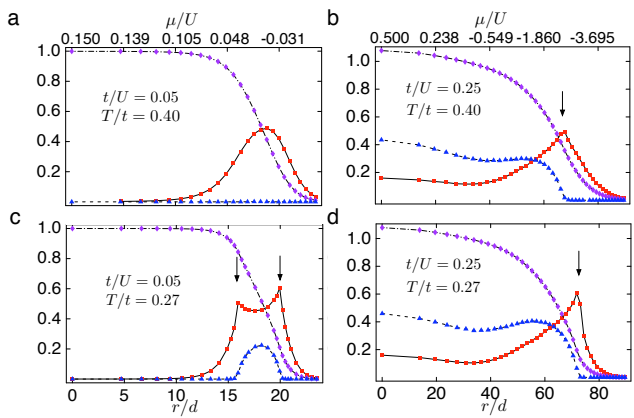}
\caption{Origin of kinks: The local density $\rho(\mu(r))$(purple), compressibility $\kappa^h(\mu(r))$(red) and superfluid density $\rho_s^h(\mu(r))$(blue) as functions of the radial coordinate in the trap. Panel (a) shows when $\rho_s=0$ in the trap,  $\kappa(r)$shows a smooth variation. As $T$ is reduced or $t/U$ is increased (b,c,d) and a finite $\rho_s$ develops in some portion of the trap, $\kappa(r)$ shows sharp kinks. The location of these kinks coincides with the S-N boundary in the trap.  Notice the location of the kinks shifts as the S-N boundary changes.}
\end{figure}         
       
The importance of Fig. 4 and 5 is that purely from a high resolution scan of the local density it is possible to obtain the local compressibility and from the existence of kinks deduce that there must be superfluid regions in the trap separated from normal regions, even though both of them have non-integer filling and the density changes smoothly across the phase boundary.  We would like to point out that there are significant differences between our work and previous studies that have primarily focused on the compressibility of Mott and superfluid phases at $T=0$\cite{17,18,19}. As seen from Fig. 5(a), the compressibility of the Mott state is exactly zero and becomes finite in the superfluid. At finite $T$, nothing singular happens as we move from Mott-like to more compressible regions in the trap. It is only at the S-N boundary (Fig. 5(b,c,d)) even though formed between two compressible phases, that the system shows enhanced critical fluctuations detected in the local compressibility. Note that in Fig. 5(b, d) there are kinks even in the absence of any Mott regions. 
       
Usually strong interaction effects in an optical lattice hamper the determination of the temperature $T$ and the chemical potential $\mu_0$ at the center in the optical lattice. We propose two methods: (i) By fitting $\rho(r)$ with $\rho(\mu(r),T)$ constrained by the known number of bosons in the trap according to $N=\int d^3r \rho(r)=\int_{-\infty}^{\mu_0}d\mu\frac{\sqrt{2(\mu_0-\mu)}}{(m\omega^2)^{3/2}}\rho(\mu)$. (ii) By fitting the tail of the density profile in the trap to the expected ideal gas behavior $\rho_{ideal}=\int\frac{d^3k}{(2\pi)^3}\left({e^{\beta(\epsilon_k-\mu_0+m\omega^2r^2/2)}-1}\right)^{-1}$  as demonstrated in Fig. 4(b). This is a direct method without any input from simulations.  Thus from the location $r_c$ of kinks in $\kappa_{diff}(r)$, experimentalists can determine $\mu_c=\mu_0-m\omega^2r_c^2/2$  to map out the phase diagram in the $\mu-T$ plane for a fixed $t/U$ and compare with theoretical results in Fig. 2(c).
       
Mapping the phase diagrams of strongly correlated quantum models is the central goal of condensed matter physics. In that context our proposed method of probing fluctuations of the density, specifically singularites or cusps, arising from critical fluctuations, provides a general method for identifying the phase boundaries. The cold atom emulator of a mathematical model in a confining trap, essentially generates a chemical potential scan of the phase diagram from a single measurement. Our studies open up new directions for mapping the finite temperature phase diagrams of strongly correlated quantum models, usually not available by other means, by probing local properties of trapped atoms in optical lattices rather than relying on the momentum distribution.

\end{document}